\documentclass[10pt,conference,letterpaper]{IEEEtran}
\usepackage{times,amsmath,epsfig}
\usepackage{soul, url}
\title{Signature Generation for Sensitive Information Leakage in
  Android Applications}
\author{
{Hiroki Kuzuno{\small $~^{\#1}$}, Satoshi Tonami{\small $~^{\#2}$} }%
\vspace{1.6mm}\\
\fontsize{10}{10}\selectfont\itshape
$~^{\#}$Intelligent Systems Laboratory, SECOM, Tokyo, Japan\\
\fontsize{9}{9}\selectfont\ttfamily\upshape
$~^{1}$h-kuzuno@secom.co.jp
$~^{2}$s-tonami@secom.co.jp%
}
\begin{document}
\maketitle
\begin{abstract}
In recent years, there has been rapid growth in mobile devices such as
smartphones, and a number of applications are developed specifically
for the smartphone market. 
In particular, there are many applications that are ``free'' to the user,
but depend on advertisement services for their revenue.
Such applications include an advertisement module - a library provided
by the advertisement service - that can collect a user's sensitive
information and transmit it across the network.
Such information is used for targeted advertisements, and user
behavior statistics. 
Users accept this business model, but in most cases the
applications do not require the user's acknowledgment in order to
transmit sensitive information. Therefore, such applications' behavior
becomes an invasion of privacy.
In our analysis of 1,188 Android applications' network traffic and
permissions,
93\% of the applications we analyzed connected to multiple
destinations when using the network.
61\% required a
permission combination that included both access to sensitive
information and use of networking services.
These applications have the potential to leak the user's sensitive
information.
Of the 107,859 HTTP packets from these applications, 23,309 %HTTP
(22\%) contained sensitive information, such as device identification
number and carrier name.
In an effort to enable users to control the transmission of their
private information,
we propose a system which, using a novel clustering method
based on the HTTP packet destination and content distances, generates
signatures from the clustering result and uses them to detect
sensitive information leakage from Android applications.
Our system does not require an Android framework modification or
any special privileges.  Thus users can easily introduce our system to their
devices, and manage
suspicious applications' network behavior in a fine grained manner.
Our 
system accurately detected 94\% of the sensitive information leakage
from the applications evaluated and produced only 5\% false negative
results, and less than 3\%
false positive results.
\end{abstract}

% NOTE keywords are not used for conference papers so do not populate them
\section{Introduction} \label{section:introduction}
With the increasing popularity of smartphones and tablets, development
for mobile device operating systems (particularly for Apple's iOS and
Google's Android, which are the most popular choices) has drastically
increased, especially the development of applications for sale in the
providers online marketplaces such as the AppStore and Google Play,
respectively.
In May 2012, Google Play alone had 500,000 applications.
Applications are categorized as free or paid. In this work, we are
primarily interested in free applications, since these often come with an 
advertising module.

A smartphone retains various kinds of personal information, such as
the contents of the user's address book, location tracking data, and
the unique device identifier.
In order to decouple the features of the device (ie, network access,
the camera, the previously discussed sensitive information), and thus
maintain security,
Android provides a framework which requires applications to have
specific permissions for accessing restricted resources.
However, the Android permission framework does not completely protect
the user's sensitive information:
applications or advertising modules send the user's sensitive
information to the outside servers using the network
\cite{hornyack11ccs, enck11usenix, grace12wisec, stevens12most}.
While this information is generally used for targeted advertising, it
can also be discovered and used by malicious parties without the
original user's awareness to combat this threat \cite{karelog,
  karelog11news}.
Various methods, employing tracking information flow and privilege
separation,
have been examined
\cite{enck10taintdroid, felt11usenix, jeon12drandroid,
  pearce12asiaccs, shekhar12usenix}.
These approaches could effectively expose a sensitive information
leakage, but they all would require extensive modifications to the
Android framework.
We think it is sufficient to force applications to notify users of
information usage details,
thus enabling the user to dynamically control the handling of
sensitive information.
Our goal is a practical method that can identify a sensitive
information leakage in applications without an Android framework
modification.
A user can easily add our system to his device and use it to manage
the transmission of sensitive information by applications. In this
way, the user can reduce the possible violations of his privacy.

In this paper, we present a novel method of clustering that uses
selected HTTP packets to generate signatures which can accurately
identify new HTTP packets that contain sensitive information.
Our primary concern is not malware, but free applications
which risk leaking sensitive information.
In many cases, malware is detected by the anti-virus software. If free
software is not malware but causes a sensitive information leakage. We
postulate that software is the new threat for the user.
In our evaluation, we analyzed 1,188 free Android applications from
the Top 100 list and the Recent Uploads list in Google Play Japan and
collected 107,859 HTTP packets that these applications generated.
We examined the number of HTTP packets that included sensitive
information and sent it to outside servers. Of our trace, 23,309 HTTP
packets 
contained such information. 
We then applied clustering to a sample of the refined data the packets
containing sensitive information to generate signatures, and
re-applied these signatures to the entire dataset.
This method result in a high percentage of true positives, and a low
percentage false positives.
Thus, we conclude that our generated signatures have sufficient
accuracy 
for detecting of sensitive information
transmission in applications.

We consider the main contributions of this paper to be:

\begin{itemize}
\item a novel clustering method using HTTP packet distance that
  identifies the similarity between the two of Android application
  network packets 
\item a system, using that the proposed method followed by signature
  generation, that can detect sensitive information leakage without
  altering the Android framework.
\end{itemize}

The rest of this paper is organized as follows. In Section
\ref{section:background}, we explain the Android architecture and
permission framework.  We describe the Android application permission
pattern can be a problem and show the practical validity of this
problem with an analysis of applications network behavior in Section
\ref{section:problem}.  We present our algorithms for the HTTP packet
clustering and signature generation in Section
\ref{section:approach}. We evaluate our approach using an HTTP packet
dataset in Section \ref{section:experiments}. 
We review related work, discuss our results, and the limitations of
this approach in Section \ref{section:relatedwork_discussion}
Finally, we conclude and suggest directions for future
work in Section 8.

\section{Background} \label{section:background}

\subsection{Android Architecture}
\begin{figure}[tp]
  \begin{center}
    \epsfig{file=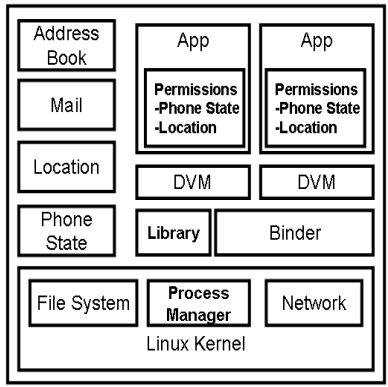, height=50.88mm, width=50.88mm}
  \end{center}
  \caption{Overview of Android architecture}
  \label{fig:android_architecture}
\end{figure}

Figure \ref{fig:android_architecture} shows an overview of Android
architecture. Android consists 
a Linux kernel, Middleware, Android applications, and sensitive
information (Address Book, GPS, Mail, Phone State are
shown). Middleware includes the Binder, the Library framework, and one
Dalvik Virtual Machine (DVM) per application.

The Linux kernel provides some fundamental features for the upper
layers: process management, file system and network services.
Middleware provides DVMs, which are used run applications as well as
the Binder, which supports IPC and checks an application's permission
list when it tries to access sensitive information via the Library.
Applications on Android have a unique Linux UID and the associated
permissions. This environment paradigm is called sand-boxing.  The
application can only
access resources within the bounds of its privileges.

\subsection{Android Permissions}
Android provides the permission framework for managing of
an application's privileges.
In order to access resources on Android, an application needs a
specific set of permissions which link to the resources.
For instance, the {\bf INTERNET} permission can connect to any outside
server using network.
The {\bf READ\_PHONE\_STATE} permission can get the unique device
identifier and line number on the device.
At the present time there are 125 privileges permissions defined  
by Android API Level 15 \cite{androidpermission}.
When an application accesses a controlled resource object, the Binder
takes charge of the reference monitor to manage the application's
request.
The Binder verifies that the application has the appropriate
permissions to bind to the requested resource.

\section{Problem Description} \label{section:problem}
In this section, we explain 
how particular combinations of application permissions can allow a
violation of user privacy.
Then, we use our analysis of the network traffic of 1,188 free
applications - how many servers are connected to by an application;
what, if any, sensitive information is included in the traffic - to
show that this problem is a practical concern.

\subsection{Application Request Permissions}
Previous studies show that many applications require the {\bf INTERNET}
permission \cite{cbarrera10ccs}.
\begin{table*}[tc]
\centering
\caption{{\it Number of applications with dangerous permission
    combinations.} Out of 1,188 applications total, 61\% (the four
  lower rows in this table) required both the {\bf INTERNET} and
  atleaset one permission for sensitive information.}
\begin{tabular}{c|c|c|c|c}
\hline
\hline
{\bf INTERNET} & {\bf LOCATION} & {\bf PHONE\_STATE} & {\bf CONTACTS}  & {\bf \# Apps}\\ \hline
x & & & & 302\\
x & & x & & 329\\
x & x & x & & 153\\
x & x & & & 148\\
x & x & x & x & 23\\
\hline
\end{tabular}
\label{tb:application_number_of_permissions_combination}
\end{table*}
Table \ref{tb:application_number_of_permissions_combination} shows the
permissions held by our collected 1,188 applications.
302 applications (25\%) require only the {\bf INTERNET} permission,  while
727 applications (61\%) require 
the {\bf INTERNET} and some combination of sensitive information
permissions. We consider sensitive information permissions to include
{\bf LOCATION},
{\bf READ\_PHONE\_STATE}, and {\bf READ\_CONTACTS}.
Those 727 applications can access sensitive resources on the device
and 
send information gathered from those sensitive resources using
the network feature, all without user confirmation,
putting the user's privacy at risk.

In Android's current model, an application requests permissions once,
on installation.  Once the application is installed, all its
transmissions are opaque to the user, who has no way of determining if
sensitive information is present in his network traffic. He may wish
to use an application without interruption when it is only
transmitting benign data, but to be prompted for confirmation when the
application wishes to send sensitive information over the network.

\subsection{Application Traffic Analysis}
It has been shown that some applications transmit sensitive
information to external servers \cite{hornyack11ccs, enck11usenix}.
One of the main reasons for this is that developers build an
advertisement module into the free version of their applications for revenue.
In order to collect statistical information of the device usage and to
provide a targeted advertisements for users, advertisement modules
take advantage of their ability to access sensitive information.

Unique device identifiers (UDIDs) are most commonly used
by  advertisement modules \cite{grace12wisec}.
The types of UDIDs include the Android ID, the International Mobile
Equipment Identity (IMEI), the International Mobile Subscriber
Identity (IMSI) and the SIM Serial ID.  Additionally, some modules
compute UDID's hash with a cryptographic hash function at the time of
transmission \cite{stevens12most}.
These UDIDs are immutable and linked to a user's real name and bank account.
Unlike Internet cookies and IP addresses, UDIDs are hard (if not impossible) to
change or erase, so it can be very dangerous for a user to have an advertisement 
module leaking his UDIDs.
If an advertisement module generates a UDID's hash value from only a
UDID,
the hash value is same all the time,
thus the user cannot change the UDID's hash value without changing the
original UDID, so the hash values have similar security problems.  We
believe that an advertisement module should use an application's
unique user ID value (i.e UUID value) rather than its UDID.
If UUIDs (which are mutable) were used instead of UDIDs, harvested
information would be restricted to the transmitting application, and
the user would have the ability to alter his device's ID if he were
concerned by the accumulation of information.

\begin{table*}[tc]
\centering
\caption{{\it HTTP packet destinations.} This table shows the number of packets
  sent to each HTTP Host destination, and the number of applications that
  send packets to each HTTP Host destination.}
\begin{tabular}{l|c|c}
\hline
\hline
{\bf HTTP Host Destination} & {\bf \# Packets}  & {\bf \# Apps}\\ 
\hline
doubleclick.net & 5786 & 407\\
admob.com & 1299 & 401 \\
google-analytics.com & 3098 & 353\\
gstatic.com & 1387 & 333\\
google.com & 3604 & 308 \\
yahoo.co.jp & 1756 & 287\\
ggpht.com & 940 & 281\\
googlesyndication.com & 938 & 244\\
ad-maker.info & 3391 & 195\\
nend.net & 1368 & 192\\
mydas.mobi & 332 & 164\\
amoad.com & 583 & 116\\
flurry.com & 335 & 119\\
microad.jp & 868 & 103\\
adwhirl.com & 548 & 102\\
i-mobile.co.jp & 3729 & 100\\
adlantis.jp & 237 & 98 \\
naver.jp & 3390 & 82\\
adimg.net & 315 & 72\\
mbga.jp & 1048 & 63\\
rakuten.co.jp & 502 & 56\\
fc2.com & 163 & 52\\
medibaad.com & 1162 & 49\\
mediba.jp & 427 & 48\\
mobclix.com & 260 & 48\\
gree.jp & 228 & 45\\
\hline
\end{tabular}
\label{tb:number_of_application_packet_destination}
\end{table*}

\begin{figure}[tc]
 \begin{center}
  \includegraphics[scale=.34]{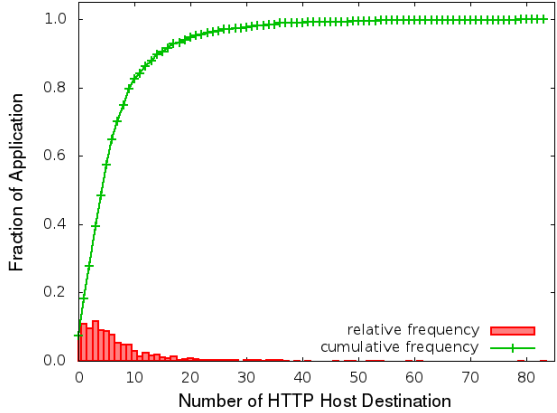}
 \end{center}
 \caption{{\it Frequency Distribution of HTTP Host Destinations.} Out
   of 1,188 applications total, 81 (7\%) have 1 destinations, 885
   (74\%) have up to 10 destinations, and average number of
   destinations was 7.9.}
 \label{fig:relative_http_host_application}
\end{figure}

\begin{table*}[t]
\centering
\caption{{\it Sensitive Information.}  This table shows for each type
  of information considered sensitive the number of packets containing
  the information, the number of application that send those
  packets, and the number of destinations to which those packets go.
}
\begin{tabular}{l |c|c |c }
\hline
\hline
{\bf Sensitive Information} & {\bf \# Packets} & {\bf \# Apps} & {\bf \# HTTP Host Destinations} \\ \hline
ANDROID ID & 7590 & 21 & 75\\
ANDROID ID MD5 & 10058 & 433 & 21\\
ANDROID ID SHA1 & 1247 & 47 & 12\\
CARRIER & 2095 & 135 & 44\\
IMEI (Device ID) & 3331 & 171 & 94\\
IMEI MD5 & 692 & 59 & 15\\
IMEI SHA1 & 1062  & 51 & 13\\
IMSI (Subscriber ID) & 655 & 16 & 22\\
SIM Serial ID & 369 & 13 & 18\\
\hline
\end{tabular}
\label{tb:number_of_application_packet_sensitive}
\end{table*}

We selected 1,188 free applications from the popular ranking in
Japan's Google Play from January to April, 2012.
Since many users choose their applications from this ranking, we
consider this to be a good sample of applications used in Japan.
We investigated the network traffic generated by these
applications. The applications sent 107,859 GET/POST HTTP packets.
The Experiment environment was a Galaxy Nexus S, Android 2.3.4.

Table \ref{tb:number_of_application_packet_destination} shows the
number of HTTP packets destined for the most common hosts and the
number of applications that send to each destination domain. Note that
many applications send HTTP packets to the same destinations,
and that some of these domains, such as ``admob.com" and ``ad-maker.info",
are clearly advertisement services. Other domains are Web API service
providers.
We can see that many of our applications send information to
advertisement servers.
We noticed during this experiment that several applications have
multiple advertisement modules (ie. AdMob, AdMaker, Adlantis, and
MicroAd).
We suspect that those applications switch from one module to another, depending on
the user's device environment such as country or carrier, to improve the
revenue.

Figure \ref{fig:relative_http_host_application} shows the cumulative
frequency distribution of HTTP host destinations of our
applications. From this, we can confirm that most of the targeted
applications connect to multiple servers.
In our examination of the HTTP host
destinations, we found that 
81 applications (7\%) have 1 destination, 885 applications (74\%) have
up to 10 destinations, and 1,006 (90\%) application have up to 16
destinations. The average number of destinations was 7.9.
One application included an embedded browser, and thus had the largest
number of destinations at 84.

Table \ref{tb:number_of_application_packet_sensitive} shows the number
of HTTP packets, applications, and HTTP host destinations that are touched 
by sensitive information, where sensitive information is considered to be: UDIDs
(IMEI, IMSI, SIM Serial ID, and Android ID), UDIDs hashed values, and
CARRIER names. IMEI refers to the assigned device number, IMSI to the
assigned telephone service subscriber number in the SIM card, SIM
Serial ID to the assigned SIM card number, and the Android ID to the assigned
Android instance number, which is generated at Android's initial boot.
The Android ID is the most frequently used identifier. 
We also found many examples of sensitive information being sent to the same
destination.
For example: ``ad-maker.info", ``mydas.mobi", ``medibaad.com", and
``adlantis.jp" expect IMEI and Android ID; ``zqapk.com" expects IMEI,
and SIM Serial ID, and Carrier name; and ``googlesyndication.com" and
``admob.com" expect only Android ID.

From these results, we can see that the user's sensitive information
is accessed by applications, which send it to outside servers via the
network.
Since Android does not provide the usage history of runtime
applications' permissions, the users can not observe the application's
network behavior, and thus can not prevent the sensitive information
leakage.

\section{Approach} \label{section:approach}
We present the following HTTP packet clustering and signature
generation methods to address the problem described in Section
\ref{section:problem}.
The objective of our work is to, without an Android framework
modification, detect suspicious network behavior specifically the
transmission of sensitive information by an application to an outside
server.
Additionally, our system should be practical and lightweight for users
to apply, and should not require any special device privileges.
Ideally, users would install our application component to handle all
the network transmissions generated by other applications.

Our approach is to collect network traffic
and generate signatures from the clustering of the traffic.
If sensitive information is sent unencrypted over the network, it is a
fairly simple matter to detect such transmission. However,
the signature generation can help to counteract leakage in polymorphic
and obfuscation traffic.
It is also effective against encrypted traffic that uses the same
encryption key over a variety of modules or applies a cryptographic
hash function to sensitive information.

\subsection{Overview}
\begin{figure}[tc]
 \begin{center}
  \includegraphics[scale=.35]{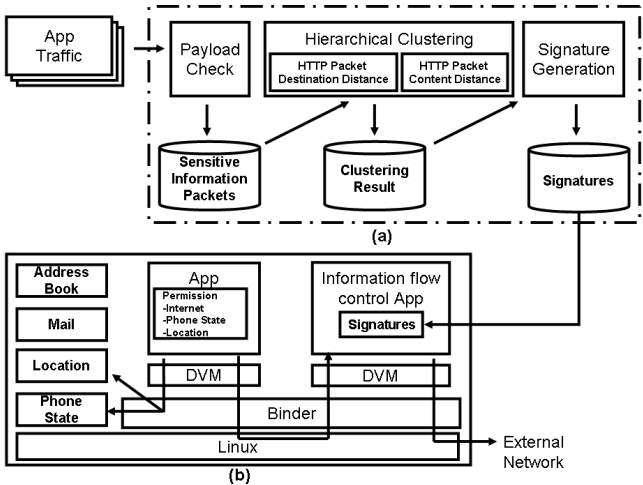}
 \end{center}
 \caption{(a) The architecture of our clustering and signature generation
   system. (b) The information flow control application that uses
   the signatures generated by (a).}
 \label{fig:system_architecture}
\end{figure}

Figure \ref{fig:system_architecture} shows our approach,
which consists of two parts. First, a separate server (shown in Figure
\ref{fig:system_architecture}a) collects application traffic,
clustering the data and generating signatures. Second, an information
flow control application on the user's device (shown in Figure
\ref{fig:system_architecture}b) fetches signatures from the servers
and manages the transmission of other applications' network traffic.

The server generates signatures by the following process.
First, it generates a payload check, which separates application network traffic into
two groups: 
one containing packets with sensitive information, and the other not.
Second, the server clusters the group containing sensitive information
based on packet destination distance and contents distance, and
constructs a set of signatures using conjunction signatures
\cite{newsome05polygraph}.
This process is most effective with accurate patterns of sensitive information
leakage, and our clustering and signature choices reflect that.
Using the HTTP packet distance is emphasizes patterns in HTTP
packets,
allowing us to distinguish trends and distributions of
HTTP packets.
Thus a packet with sensitive information will be clustered with other
packets containing sensitive information, generating a useful
signature.  In order to generate such useful signatures, we define
distance to include both packet content and packet destination. This
broader definition causes results sent to the same server to be
clustered together, creating advertisement module specific signatures.
The information flow control application inspects network traffic
using the Android API and detects sensitive information leakage using
the our server generated signatures.
It does not require any special privileges.

\subsection{HTTP Packet Destination Distance} \label{subsection:http_packet_destination_distance}
The HTTP packet destination distances are calculated by the packets'
destination IP addresses, port numbers, and HTTP host domains.
Given two HTTP packets
$p_{x}$ and $p_{y}$, we define the HTTP packet destination distance as
\begin{equation*}
d_{dst}(p_{x}, p_{y}) = d_{ip}(p_{x}, p_{y}) + d_{port}(p_{x}, p_{y}) + d_{host}(p_{x}, p_{y}).
\end{equation*} 
Let HTTP packet $p_{n}$ destination be defined as $p_{n} =\\ \{ip_{n},
port_{n}, host_{n}\}$, where $ip_{n}$ is a destination IPv4 address,
$port_{n}$ is the port number, $host_{n}$ is HTTP host. The distance
in the above equation are defined as follows:
\begin{itemize}
\item Destination IP Address Distance: The distance between
  destination IP addresses' high bit is the longest matching prefix of
  the binary representations. IPv4 addresses have a $2^{32}$ bit
  space, and IP address blocks are denoted approximately by the upper 8
  bit range. IP address blocks are allocated to organizations by the
  National Internet Registry and if the upper bits of IP addresses
  match on separate packets, there is a high possibility that the two
  destinations are managed by the same organization.  Therefore, we
  define the destination IP address distance on packets $p_{x}$,
  $p_{y}$ as
\begin{equation*}
d_{ip}(p_{x},p_{y}) = lmatch(ip_{x}, ip_{y}) / 32 \in [0,1]
\end{equation*}
where $lmatch$ is a function returns a number of common upper bits in
two IP address.
\item Port Number Distance: The distance between port numbers is a
  Boolean (matching or not). Port numbers have a $2^{16}$ bit space,
  and usually, specific port number is reserved for services. We
  define the port number distance on packets $p_{x}$, $p_{y}$ as
\begin{equation*}
d_{port}(p_{x}, p_{y}) = match(port_{x}, port_{y}) \in \{0,1\}
\end{equation*}
where $match$ is a function returns 1 on matching ports, and 0 on different ports.
\item HTTP Host Distance: We define the HTTP host as the character
  string of the FQDN. Thus, the distance between HTTP host domains 
  can be computed using the generality method to determine their edit
  distance.
  We define the HTTP Host distance
  on packets $p_{x}$, $p_{y}$ as
\begin{equation*}
d_{host}(p_{x}, p_{y}) = \frac{ed(host_{x},
  host_{y})}{max(len(host_{x}), len(host_{y}))} \in [0, 1]
\end{equation*}
where $ed$ is a function which returns an edit distance result,  $len$
is a function which returns a length of character strings, and  $max$ is a
function which returns the greater of its two input values.
\end{itemize}

\subsection{HTTP Packet Content Distance} \label{subsection:http_packet_content_distance}
The HTTP packet content distance is computed using the request-line,
cookie, and message-body fields of the HTTP header. 
Given two HTTP packets $p_{x}$ and $p_{y}$, we define the HTTP content
distance $d_{header}(p_{x},p_{y})$ as
\begin{equation*}
d_{header}(p_{x}, p_{y}) = d_{rline}(p_{x}, p_{y})
+d_{cookie}(p_{x}, p_{y}) + d_{body}(p_{x}, p_{y}).
\end{equation*}
Let HTTP packet $p_{n}$ contents be defined as $p_{n}
=\{rline_{n},\\ cookie_{n}, body_{n}\}$, where $rline_{n}$ is
request-line, $cookie_{n}$ is cookie, $body_{n}$ is message-body.
These contents are character or binary strings. In order to accurately
compute a distance, we apply the normalized compression distance (NCD)
\cite{IEEEexample:ncd} algorithm, which is based on Kolmogorov's
complexity, to calculate the closeness of two strings without any
context dependency.
The NCD of any two character strings is defined as
\begin{eqnarray*}
ncd(x,y) = \frac{C(xy)-min(C(x), C(y))}{max(C(x),C(y))}
\end{eqnarray*}
where $C(x)$ is a function which compresses a character string $x$, then
returns its length. We define the distance between content components
of HTTP packets $p_{x}$, $p_{y}$ as
\begin{eqnarray*}
d_{data}(p_{x}, p_{y}) = ncd(data_{x}, data_{y}) \in [0,1]
\end{eqnarray*}
where $data$ corresponds to request-line, cookie, and message-body respectively.
After each $data$ has been computed, they are combined into the overall distance.

\subsection{Hierarchical Clustering} \label{subsection:hierarchical_clustering}
Hierarchical clustering uses group averages for iterative calculation
and computes the proximity of clusters with HTTP packet distance (HTTP
packet destination distance and HTTP packet content distance) as a
heuristic.  It then assigns a cluster to each HTTP packet, and
iteratively composes new clusters from the nearest distance of HTTP
packet pairs until there is only one cluster.
Given two HTTP packets $p_{x}$ and $p_{y}$, we define the HTTP packet
distance as
\begin{eqnarray*}
d_{pkt}(p_{x}, p_{y}) = d_{dst}(p_{x}, p_{y}) +
d_{header}(p_{x},p_{y})
\end{eqnarray*}
using the formulae from sections \ref{subsection:http_packet_destination_distance}, 
\ref{subsection:http_packet_content_distance}
to compute $d_{dst}$ and $d_{header}$. Given two clusters $C_{x}$ and
$C_{y}$, group average distance is defined as
\begin{eqnarray*}
d_{group}(C_{x}, C_{y}) = \frac{1}{|C_{x}||C_{y}|} \sum_{p_{x}\in
  C_{x}} \sum_{p_{y}\in C_{y}}d_{pkt}(p_{x},p_{y}).
\end{eqnarray*}
In a dataset of $N$ HTTP packets ${\bf H}=\{p_{i}\}_{i=1..N}$, we
apply hierarchical clustering to 
a subset ${\bf P}$ of size $M$: ${\bf P}_{j,j=1..M} \subset {\bf
  H}$, using the following method:

\begin{enumerate}
\item Assign each HTTP packet $p_{k} \in {\bf P}$ to cluster $C_{k}$.
  At the end of this step, ${\bf C}=\{C_{k}\}_{k=1..M}$ is the set of
  defined clusters.
\item Chose any cluster $C_{x}\in {\bf C}$, and compute the distance
  to all other clusters $C_{y, y=1..M} \in {\bf C}, x \neq y$ using
  the cluster distance $d_{group}$.
\item Select the cluster $C_{y}$ that is the closest to
  $C_{x}$. 
  Create a new cluster $C_{z}=\{C_{x}, C_{y}\}$ and add it to
  ${\bf C}$, then remove $C_{x}$ and $C_{y}$
\item Repeat until ${\bf C}$ has one cluster.
\end{enumerate}

\subsection{Signature Generation}
We generate a conjunction signature set from the hierarchical
clustering result, which is a dendrogram of the HTTP packet group. A
conjunction signature set contains the invariant tokens that describe
the longest common substrings in the dendrogram.
Signatures each represent a feature of the cluster.
That is, they reflect sensitive information as a invariant token.
Given a dataset of $N$ HTTP packets ${\bf H}=\{p_{i}\}_{i=1..N}$ and a
subset ${\bf P}_{j,j=1..M} \subset {\bf H}$ used to generate (as
described in Section \ref{subsection:hierarchical_clustering})
dendrogram ${\bf C}$, which has the nesting structure characteristic
of clusters, we generate the conjunction signature set using the
following process:
\begin{enumerate}
\item Select the top of cluster $C_{i} \in {\bf C}$.
\item Compute a signature $S_{i}$ as longest common strings
  of HTTP contents in $C_{i}$.
\item Remove $C_{i}$ from ${\bf C}$ and repeat for all
  clusters in ${\bf C}$.
\end{enumerate}

\section{Evaluation} \label{section:experiments}

\subsection{Experimental Setup}
We collected network traffic from 1,188 free applications running on
an Android 2.3.6, Galaxy Nexus S, from January to April, 2012. The
application set was as previously described in Section
\ref{section:problem}.
Each application was run manually for 5 to 15 minutes on the device.
We attempted to test a every possible application function.  We
generated the data manually, since it is difficult to automatically
test an application that requirs user interaction
such as entering passwords and other user identification, or correct
screen taps for a game.

The resulting dataset of application network traffic contained 107,859
GET/POST HTTP packets.
For this experiment, we manually separated the dataset into a
suspicious group and a normal group for the evaluation of our
signatures' detection rate.
The suspicious group consisted of packets containing sensitive
information. The normal group was made up of those containing
non-sensitive information.
Again we considered UDIDs (Android ID, IMEI, IMSI, and SIM Serial ID),
hashed UDIDs (MD5, SHA1), and Carrier names to be sensitive
information.

In this experiment, we were not concerned with encrypted packets and
obfuscation packets except the hashes mentioned above. Consequently,
the normal group contained these packets.
The suspicions group consisted of 23,309 HTTP packets, and the normal
group contained 84,550 HTTP packets. The details of the suspicious
group are shown in Table
\ref{tb:number_of_application_packet_sensitive}.
We selected $N$ HTTP packets at random out of the suspicious group for
signature generation, where $N$ was increased from 0 up to 500 in
intervals of 100. 
Finally, we applied the generated signatures to the dataset in its
entirely to see how accurately they could identify packets containing
sensitive information.
We evaluate signatures for accuracy detection rate.

\subsection{Experimental Results}
\begin{figure}[tc]
 \begin{center}
  \includegraphics[scale=.38]{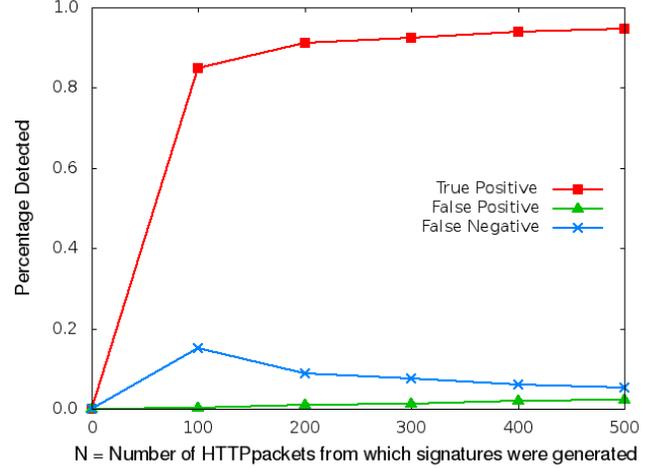}
 \end{center}
 \caption{Detection Rate of Sensitive Information Leakage}
 \label{fig:detection_rate}
\end{figure}

Figure \ref{fig:detection_rate} shows the results of our experiment.
approach. We evaluated the percentage of true positives, false
positives and false negatives for varying values of $N$.

{\bf True Positive:}
a correctly detected packet containing sensitive information. The
percentage of true positives was calculated according to the following
equation:
\begin{equation*}
{\rm TP} = \frac{{\rm \#\; of\; detected\: sensitive\; information\;
    packets} - N} {{\rm \#\; of\; sensitive\; information\; packets} - N}
\end{equation*}
There were 23,309 sensitive information packets in the dataset for our
evaluation. 
Our system produced 85\% true positives at sampling size $N=100$. It
grew beyond 90\% by $N=200$, with the best result being 94\% at $N=500$.
These results show that true positives rise with an increasing number
of signature generating sensitive information packets, therefore,
signatures generated from more packets cover a wider common pattern of
information leakage.

{\bf False Negative:}
a sensitive packet that was not correctly detected. We calculated the
percentage of false negative results using following equation:
\begin{equation*}
{\rm FN} = \frac{{\rm \#\; of\; un detected\: sensitive\; information\;
    packets}} {{\rm \#\; of\; sensitive\; information\; packets} - N}
\end{equation*}
As stated above, there were 23,309 sensitive packets in our dataset.
In this experiment, there were 15\% false negatives at $N=100$, only
8\% or less at more than $N=200$, and finally 5\% at $N=500$.  Thus,
effective detection of information leakage is improved by increasing
the number of sensitive information packets used for generating
signatures.

{\bf False Positive:}
a non-sensitive packet incorrectly detected as sensitive.
We calculated the percentage of false positives using following equation:
\begin{equation*}
{\rm FP} = \frac{{\rm \#\; of\; detected\: non\mathchar"712D sensitive\;
    information\; packets}} {{\rm \#\; of\; non\mathchar"712D sensitive\;
    information\; packets} - N}
\end{equation*}
This value is important for an evaluation of our system's signatures
detection rate in terms of usefulness.
If our system produces many false positives, users will be continually
bothered by unnecessary warnings and prompts.
This dataset had 84,550 non-sensitive information packets. The
signatures from this dataset produced 0.3\% false positives at
$N=100$, 0.9\% at $N=200$, and eventually 2.3\% at $N=500$. 
We can see that verbose signatures are generated by increasing the
number of clustering packets. 
We postulate that large signature generating sets produce signatures
that detect packets without relation to their information leakage.

\section{Related Work and Discussion} \label{section:relatedwork_discussion}
Other approaches to preventing sensitive information leakage include
taint tracking and permission framework modifications.  In this
section, we compare our approach with current research results, and
discuss the limitations of our scheme.

Several studies have analyzed the security and privacy concerns of the
potential sensitive information leakage in Android and iOS
applications \cite{hornyack11ccs, enck11usenix, schlegel11ndss,
  gilbert11mcs}.
Other works have focused specifically on advertisement modules' access
to device identification number and location information and their
ability to send it over the network \cite{grace12wisec,
  stevens12most}.
To address this problem, fine-grained access control techniques have
been proposed. These projects implement enhancements to the Android
permission policy \cite{enck09ccs, ongtang09acsac, nauman10apex}.
Taint tracking can also accurately detect sensitive information
leakage and control the information flow of applications
\cite{hornyack11ccs, enck10taintdroid}. These approaches have shown
that dynamic analysis of the trace details of applications' behavior
on the Android framework can ameliorate the problem. It has the
advantage of having low overhead with very few false positives,
minimizing the notifications issue to the users.
Separating the advertisement module's permissions from application's
can also reduce the privacy risk \cite{pearce12asiaccs,
  shekhar12usenix}. If the user does not grant permissions to the
advertisement module, he can be sure that advertisement module does
not access sensitive information on the device and send it over the
network.
However, all these approaches require Android framework
modifications. Our approach of generating signatures that can identify
sensitive packets with a small percentage of false positives does not
require any modifications to the Android framework, or any escalation
of user privilege on the Android device, making our system simple and
immediately applicable.  It can also be used with previous works
should the proposed modifications be implemented or with anti-virus
applications which are designed to detect malware.

Generating signatures from clustering is not a new idea. Previous work
on signature generation used clustering focused on the similarity of
network traffic or on the characteristics of applications,
specifically, targeting malware and malicious network traffic
\cite{gu08ndss, chung09ipom, ingham07raid, wehner08journal,
  bailey07raid, bayer09ndss, coull11ndss}.
Other proposals regarding clustering network destination and traffic
individually intend to comprehend some aspect of an application's
behavior \cite{gu08usenix}, and signature generation uses include
computing HTTP packet statistics
contents to improve detection rates \cite{perdisci10nsdi}, similar to
our approach.

Clustering in general is useful for pulling together patterns in large
amounts of data, but the number of the generated signatures tend to
increase with cluster size, and can produce signatures that match most
network packets (e.g {\bf POST *}, {\bf GET *}, {\bf * HTTP/1.1} ), if
the signature generation is applied carelessly.
For this reason, it has been difficult for a signatures approach to
achieve high detection rates using a real dataset.
Probabilistic signatures \cite{newsome05polygraph, li06oakland,
  kong10securecomm} might improve detection of information leakage on
Android applications, and we hope to include them in our scheme
in future work.
Currently our use of HTTP packet destination distance allows us to
generate more useful signatures.
A concern with using the destination distance is that two HTTP packets
may have close IP addresses but be owned different organizations, thus
generating an erroneously small distance is still very small.  While
our current implementation does not specifically handle this case, we
feel that using a registration information process such as WHOIS could
be helpful for the verification of IP addresses and domain names,
which could be used to confirm the distances. Our current approach
also does not focus on encrypted or obfuscated traffic. It can be
difficult
to detect sensitive information in SSL traffic but if an advertisement
module uses one encryption key among applications or applies a
cryptographic hash function to sensitive information, our approach can
detect it.

Privacy preserving advertisement approaches which do not collect
users' behavioral information from devices for targeted advertising
have also been proposed \cite{toubiana10ndss, bilenko11pets}.  In
spite of these proposals reducing users' privacy risks, they are not
being utilized practically.

\section{Conclusion}
Advertisement services are widely accepted among application
developers. However it is still important to investigate the behavior
of an application with regards to security and privacy.
We have shown that many Android applications require permissions for
sensitive information access and network features, and that among them
are applications that connect to many outside servers without the
user's acknowledgment.
Furthermore, we have observed that applications' network behavior
includes a large amount of sensitive information, particularly, UDIDs
and Android ID which are immutable identifiers.
We have proposed a novel clustering method using HTTP packet distances
which include both the distance between HTTP packet destinations and
between HTTP packet contents. 
Using that clustering method in combination with signature generation
in our dataset of 1,188 applications and 107,859 packets, which
included 23,309 sensitive information packets, we were able to achieve
94\% accurate detection of packets containing sensitive data with only
3\% false positives.

\section*{Acknowledgment}
We would like to thank Catherine Redfield
for discussions and feedback. We also thank
Hiroshi Yokoshima for help with experiments.

%The heading of the Acknowledgment section and the References section
%must not be numbered.

%Causal Productions wishes to acknowledge Michael Shell and other
%contributors for developing and maintaining the IEEE LaTeX style files
%which have been used in the preparation of this template.  To see the
%list of contributors, please refer to the top of file IEEETran.cls in
%the IEEE LaTeX distribution.

\bibliographystyle{IEEEtran}
{\footnotesize
\bibliography{moda2013_android_signature.bib}
}
\end{document}